# Triple Scoring Using Paragraph Vector

## The Gailan Triple Scorer at WSDM Cup 2017

Esraa Ali
ADAPT Centre
School of Computer Science
and Statistics, Trinity College
Dublin
Dublin, Ireland
esraa.ali@adaptcentre.ie

Annalina Caputo
ADAPT Centre
School of Computer Science
and Statistics, Trinity College
Dublin
Dublin, Ireland
annalina.caputo@adaptcentre.ie

Séamus Lawless
ADAPT Centre
School of Computer Science
and Statistics, Trinity College
Dublin
Dublin, Ireland
seamus.lawless@adaptcentre.ie

## ABSTRACT

In this paper we describe our solution to the WSDM Cup 2017 Triple Scoring task. Our approach generates a relevance score based on the textual description of the triple's subject and value (Object). It measures how similar (related) the text description of the subject is to the text description of its values. The generated similarity score can then be used to rank the multiple values associated with this subject. We utilize the Paragraph Vector algorithm to represent the unstructured text into fixed length vectors. The fixed length representation is then employed to calculate the similarity (relevance) score between the subject and its multiple values. Our experimental results have shown that the suggested approach is promising and suitable to solve this problem.

## 1. INTRODUCTION

The task of ranking multiple values of a subject property is formally known as *triple scoring*. Given a property that contains multiple values, the task is concerned with ranking the values according to their relevance to a specific subject. In that context, a relevance score is usually computed to order the different values for the same property.

One possible use case for this task is the ordering of the multiple professions of an individual. For example, Neil Young is a known activist, actor, record producer, screenwriter, film score composer, film director, songwriter and film producer. This research area studies which professions are most strongly relevant to Neil Young. It investigates the capability of a system at capturing the human perception of topics, such as: Is Neil Young more likely to be a film producer than a songwriter? Is he more likely to be an actor rather than a screenwriter? Solving this problem is useful for several information retrieval tasks [1].

In this research we introduce a proposed approach to triple scoring that is based on the text describing the subject. We exploit the structure of the knowledge base that, for a given property, can present either subjects associated with multiple values but also ones that are uniquely associated to a single value. We investigate the use of the text associated with single value subjects to rank multiple value subjects.

For example, let us assume we have a group of persons (subjects) with actor (value) as a single profession (property). We can use the text describing or mentioning these persons to build a single model for the 'Actor' profession. Hence, when we want to rank the professions for a given person, with actor being one of them, we can use the 'Actor' model to compute the extent to which the text describing the person is *similar* to the text which "usually" describes any person with 'Actor' as their profession.

Given enough data, we can build a model for every possible value (profession). Then, we can compute the similarity between each of the multiple values and the description text associated with every subject. This similarity can be considered as a relevance score and used to order the relevance of multiple values to a subject.

In order to build a model for each value –assuming that all values are known in advance– and compute the similarity score, a representation for the text is needed.

In the case of the WSDM Cup 2017 Triple Score task, this does not represent a problem, since all values (200 professions and 100 nationalities) are given by the organizers and known in advance.

The problem falls within the research areas of text representation, textual similarity and topic modeling. Many statistical methods have been commonly used for this kind of text representation problem.

Recently, Paragraph Vector [5] has been demonstrated to outperform other traditional statistical methods, such as Bag-of-words and supervised Latent Dirichlet Allocation (sLDA) [6], while efficiently maintaining the semantics of the text.

Paragraph Vector is an unsupervised algorithm that learns a vector representation for whole pieces of unstructured text by using artificial neural networks (ANN). This is in contrast to other algorithms, like LSA or word2vect, which represent the semantic of single words through vectors.

Given the vector representation, we can derive the similarity score between the subject's related text and the value related text (learnt from subjects that have this as a unique value for the same property). In other words, the algorithm encodes unstructured texts in fixed-length vectors. The vectors capture the semantics of the text and their fixed length nature makes them suitable for calculating similarities.

This approach was developed in order to participate in the WSDM Cup 2017 Triple Scoring task [2, 4]. The task provided two different datasets for the evaluation of the ranking of professions and nationalities associated with the given instances. Throughout the paper we use the profession use case as an example, while in the evaluation section we present the results for both the profession and nationality datasets. Further information about the Triple Scoring Task, its motivations, related work, baseline and the description of the datasets can be found in [1].

The rest of the paper is organized as follows: the next section is dedicated to introducing the Paragraph Vector approach and its application to the Triple Scoring Task. Section 3 details the proposed methodology. Section 4 provides an overview of the results of the evaluation task. We conclude with our findings and possible enhancements in the concluding section.

## 2. BACKGROUND

In the recent few years, word2vec [7] has established itself in the area of Natural Language Processing (NLP) as a compact and efficient way for representing the semantics of words into fixed length vectors. However, the representation of complex structures, like sentences or paragraphs, is obtained through a simple algebraic operator: the addition of word vectors appearing in the text. Similarly to what happens in classical Bag-of-Word representations, this approach to text representation neglects the rules and the structure that combines words together in a text.

To this model, "hot dog" and "dog hot" have identical meaning. To this end, Paragraph Vector algorithms, originally introduced by Le et al. [5], have been introduced as a way to extend the same approach to longer sequence of text, such as sentences or paragraphs.

Paragraph Vector is an unsupervised algorithm that learns continuous distributed vector representations for pieces of text. The texts can be of variable length, ranging from sentences to documents. The model trains vectors to predict the next word in a paragraph context. It concatenates several word vectors within the paragraph into a single vector. Both Word Vectors and Paragraph Vectors are trained by stochastic gradient descent and back-propagation. In the resulting trained model, semantically similar paragraphs have similar vector representations.

In our experiments, we explored two different implementations of Paragraph Vectors. The first is the Distributed Memory model (PV-DM), with concatenation or with average. In this model the paragraph vector and word vectors are averaged or concatenated to predict the next word in a context.

The second implementation is the Distributed Bag-of-Words (PV-DBOW) method. The previous methods considers both the paragraph vector and the word vectors to predict the next word in a text window. The PV-DBOW method does not need word vectors and does not consider word ordering. At each learning step, it selects a random word from the paragraph and uses it to form a prediction task, and based on the error updates the output paragraph vector.

After being trained, the paragraph vectors can be used as features representing a paragraph or a piece of text. We can feed these features directly into various machine learning techniques, such as logistic regression, SVMs or decision trees [3].

## 3. APPROACH

The intuition of our approach is to use the text describing a subject to infer the relevance of a property value. In our use case this will be computed by using the texts mentioning a person's name to infer his profession or nationality relevance. Different methods can be followed to achieve this target. We have implemented and compared two methods. The first one relies on a Logistic Regression classifier to compute the relevance score (LogReg). The second method is based on a cosine similarity score (CosSim). Both employ Paragraph Vector internally to generate a relevance score. The final score is then mapped to the range 0 to 7 to match the human-labeled dataset.

We start by explaining how data is preprocessed and organized in subsection 3.1. The LogReg and CosSim methods are detailed in the subsequent two subsections. Subsection 3.2 illustrates the Paragraph Vector training and its parameters. The mapping equations adopted to generate the final score are listed in subsection 3.5.

All the approaches are implemented in Python. The Paragraph Vector method is implemented using the Gensim Python library for topic modeling[1] [9]. The machine learning related methods are implemented using the Scikit-learn machine learning library in Python [8].

### 3.1 Data Preparation and Enrichment

The dataset provided by the task contains triples, extracted from Freebase, for persons and their associated professions, both single and multiple professions. The same is also provided for nationalities [1]. The dataset also includes sentences with person names extracted from Wikipedia.

As a pre-processing step, for each person we built a person file containing all the sentences from the Wikipedia dataset which contain a mention of the person in question. These sentences are then converted to lowercase and the person name is removed from the text. No stop-word removal step was applied, as the stop words themselves contribute to the context learning process.

Then, all the persons with a single profession were grouped by profession. After analysing the data and a number of trials, we found that some professions were underrepresented, if represented at all, while other celebrity-related professions, like Actor or Football player, can have more than 30,000 person files associated with them. This imbalance in the dataset made the final trained model biased towards some professions.

To solve this issue, a data enrichment step was added for professions with less than 100 files. For each missing or poorly represented profession, we used the Wikipedia search API to retrieve the 200 most relevant pages.

For the overrepresented professions, we only considered the first 5,000 files for training. Due to the previous pre-processing steps, these files are already shuffled so this constitutes a random selection.

This imbalance issue was not encountered in the nationality dataset. However, some nationalities, like USA, were associated with more than 150,000 person files. In this case, the same truncation step was applied to all those nationalities with more than 5,000 person files.

### 3.2 Paragraph Vector Training

The Paragraph Vector model is trained using all persons in the dataset. For the single profession persons, the vectors representing their text are grouped in sets to be used in the LogReg or CosSim methods later.

The vectors belonging to persons with multiple professions will be used in the final score calculation step of each method. Three Paragraph Vector implementations were experimented for training the vector representations and the best performing one was chosen for the submission. The implementations consist of: Paragraph Vector Distributed Bag-Of-Words (PV-DBOW), Paragraph Vector Distributed Memory (PV-DM) with concatenation and PV-DM with average [3]. In our training the PV-DBOW performed the best in both LogReg and CosSim methods.

The Paragraph Vector model was trained using 20 epochs and the dataset was shuffled between epochs. The window size was set up to 5 for preserving some context. The minimum count for words to be considered in the vocabulary was 10. The number of workers (concurrent running threads) was set up to 40. Negative sampling was used to remove noisy words and it was set to 5. The dimensionality of the output vectors was set to 200.

### 3.3 LogReg

In **LogReg** we train a Logistic Regression classifier to score how a property value is relevant to the subject's text. It computes the probability that a text description is associated with a single profession, like for example 'Lawyer'. The classifier is trained using text belonging to persons with a single profession. Unstructured raw text cannot be inputted directly to the classifier, instead it is converted to fixed length paragraph vectors and those vectors are used as input features to the classifier. The features used in the training are the person's paragraph vector and the label is the person's profession retrieved from the knowledge base. The trained classifier is then used to predict the probability that a person belongs to a profession. For persons with multiple professions, we use this probability to rank their professions.

---
[1] https://radimrehurek.com/gensim/models/doc2vec.html

In order to clarify the idea, let us consider the Actor profession. First, we collect all vectors for persons who have Actor as their only profession; we repeat this for all the other professions (200) that are in the dataset. The logistic regression classifier is then trained using the labeled vectors. At this step the classifier should be able to give a probability that the person X belongs to the profession Y given the input paragraph vector. In the inference stage, the vector for the person with multiple professions is fed to the trained model. The model computes the probability of the profession given the person vector as input. We finally use this probability as the relevance score.

### 3.4 CosSim

The second approach, **CosSim**, calculates the cosine similarity between the grouped set of vectors for each profession and each multiple profession person vector. Let us assume that the vectors for the Actor profession are averaged in one vector $\vec{a}$. Paragraph Vector generates for the person which has multiple professions (including actor) the vector representation $\vec{p}$. The method computes the dot product between the vectors $\vec{a}$ and $\vec{b}$.

By doing so, it measures the textual similarity between the texts mentioning a person and the text that usually describes other actors. The measure is calculated for all the professions associated with the person and then it is used as relevance score.

### 3.5 Mapping and Final Score Generation

The output score for the two experimented methods is in the range from 0 to 1. A mapping step is needed to scale the values to match the human labeled score ranging from 0 to 7. These scores were provided as part of the task. Different mapping techniques were explored during training. In particular, we experimented with the MapLog and MapLin methods presented in [1]. In addition to that, we also experimented with the MapRange method. MapRange considers the minimum and the maximum of an input array while calculating the mapped value. Given an array $A$ of prediction values, we compute MapRange as follows:

$$\text{score} = \frac{\text{maxValue} * (\text{value} - \min(A))}{\max(A) - \min(A)} \quad (1)$$

Where $maxValue$ is 7, and the $value$ is the number to be mapped. During training, we found that MapLin gives the best results.

## 4. EVALUATION RESULTS

Table 1 shows the evaluation results for the nationality data comparing the two methods LogReg and CosSim, while Table 2 reports the results for professions.

The results of all the participants in the task are reported in the WSDM Cup 2017 Triple Scoring Task overview paper [2]. The metrics used during the evaluation are: Accuracy (with delta=2), Kendall's Tau correlation and the Average Score Difference (ASD).

| Method | Accuracy | Kendall's Tau | ASD |
|--------|----------|---------------|------|
| LogReg | 0.62 | 0.36 | 2.29 |
| CosSim | 0.80 | 0.39 | 1.40 |

**Table 1: Evaluation results for nationalities datasets.**

Although it is simpler, the CosSim method outperforms the LogReg approach in both datasets with respect to all the employed metrics. The similarity calculation is less affected by the number of persons that exist for each profession. Changing the classification algorithm to SVM and normalizing the input paragraph vector values as well as tuning the logistic regression parameters had no significant improvement on the results. The classifier was always negatively affected by the unbalanced training dataset.

In other words, whenever the classifier sees a profession like Actor (which has the highest number of instances) it gives it the highest probability regardless whether it is the most important profession for an individual.

| Method | Accuracy | Kendall's Tau | ASD |
|--------|----------|---------------|------|
| LogReg | 0.63 | 0.29 | 2.30 |
| CosSim | 0.68 | 0.34 | 1.94 |

**Table 2: Evaluation results for professions dataset**

However, this imbalance in the dataset seems not to affect the CosSim measure. However, this measure fails to correctly distinguish related professions, such as actor, film producer, and director. The results shown in Table 2 for the profession dataset was obtained after the data enrichment step. It was observed that the CosSim measure constantly scores some professions to 0. After some investigation, we found that those professions had no or a very low number of persons and their text is not similar to any other professions (an example is the carpenter profession). Adding more examples for those professions increased the CosSim accuracy from 0.65 to 0.68.

In this case, the text representing each profession is similar due to the presence of words from the same domain. The same issue was not detected in the nationality dataset, since related countries have usually different text descriptions. Indeed, on this dataset the CosSim method gave better performance.

## 5. CONCLUSION AND POSSIBLE ENHANCEMENTS

In this paper we described our Paragraph Vector approach to the WSDM Cup 2017 Triple Scoring Task. Paragraph Vectors are trained to represent documents that describe each property value (in our case each profession or nationality). These vector representations are then used to measure the proximity between an instance (person) representation vector and the vector that represents the attribute to score (profession/nationality). We presented our experimental results on the test datasets. Results showed that the cosine similarity measure (CosSim) computed on the trained Paragraph Vectors provides a useful indicator for the relevance of the profession/nationality to a person.

Further enhancements can be implemented to improve the results. We believe that adding a learning algorithm at the last stage to learn/adjust the final score might improve the results. In addition to that, data enrichment by adding more learning examples (paragraphs) for each value (profession or nationality) can help to provide a more balanced dataset, which would result in better vector representations.

## Acknowledgement

The ADAPT Centre for Digital Content Technology is funded under the SFI Research Centres Programme (Grant 13/RC/2106) and is co-funded under the European Regional Development Fund.